\documentclass[
]{ceurart}

\usepackage{listings}
\usepackage{xcolor}

\usepackage{subcaption}

\usepackage{tcolorbox}
\usepackage{caption}

\usepackage{wrapfig}

\usepackage{float}

\newcommand{\hx}[1]{\noindent\textbf{#1}}
\newcommand{\hs}{\vspace{-.5\baselineskip}}

\begin{document}

\copyrightyear{2025}
\copyrightclause{Copyright for this paper by its authors.
  Use permitted under Creative Commons License Attribution 4.0
  International (CC BY 4.0).}

\conference{PMAI'25}

\title{XABPs: Towards eXplainable Autonomous Business Processes}

\author[1,2]{Peter Fettke}[%
orcid=0000-0002-0624-4431,
email=peter.fettke@dfki.de,
url=https://www.dfki.de/,
]
\address[1]{German Research Center for Artificial Intelligence (DFKI), Campus D3 2, 66123 Saarbrücken, Germany}
\address[2]{Saarland University, Campus D3 2, 66123 Saarbrücken, Germany}

\author[3]{Fabiana Fournier}[%
orcid=0000-0001-6569-1023,
email=fabiana@il.ibm.com,
url=https://research.ibm.com/people/fabiana-fournier,
]
%\address[3]{IBM Research, Israel}

\author[3]{Lior Limonad}[%
orcid=0000-0002-4784-2147,
email=liorli@il.ibm.com,
url=https://research.ibm.com/people/lior-limonad,
]
\address[3]{IBM Research, Israel}

\author[4]{Andreas Metzger}[%
orcid=0000-0002-4808-8297,
email=andreas.metzger@paluno.uni-due.de,
url=https://adaptive-systems.org/,
]
\cormark[1]
\address[4]{paluno (The Ruhr Institute for Software Technology), University of Duisburg Essen, Essen Germany}

\author[5]{Stefanie Rinderle-Ma}[%
orcid=0000-0001-5656-6108,
email=stefanie.rinderle-ma@tum.de,
url=https://www.cs.cit.tum.de/bpm/staff/rinderle-ma/,
]
\address[5]{Technical University of Munich, TUM School of Computation, Information and Technology, Garching, Germany}

\author[6]{Barbara Weber}[%
orcid=0000-0002-9421-8566,
email=barbara.weber@unisg.ch,
url=https://ics.unisg.ch/chairs/barbara-weber-software-systems-programming-and-development/,
]
\address[6]{University of St. Gallen, Rosenbergstrasse 30, 9000 St. Gallen, Switzerland}

\cortext[1]{Corresponding author.}
% \fntext[1]{These authors contributed equally.}

\begin{abstract}
Autonomous business processes (ABPs), i.e., self-executing workflows leveraging AI/ML, have the potential to improve operational efficiency, reduce errors, lower costs, improve response times, and free human workers for more strategic and creative work.
However, ABPs may raise specific concerns including decreased stakeholder trust, difficulties in debugging, hindered accountability, risk of bias, and issues with regulatory compliance. We argue for eXplainable ABPs (XABPs) to address these concerns by enabling systems to articulate their rationale. 
% Current explainable AI (XAI) techniques are limited in capturing the complexities of business process models. 
The paper outlines a systematic approach to XABPs, characterizing their forms, structuring explainability, and identifying key BPM research challenges towards XABPs.
\end{abstract}

\begin{keywords}
  Business Process Management \sep
  Autonomous Business Processes \sep
  Explainability \sep
  Agentic AI
  \end{keywords}

\maketitle

\section{Introduction}
\label{sec:intro}
% \textbf{lior's comment: In relation to the ABPMS manifesto [ref] and in general when developing systems where the actors (or at least some of them) are gradually evolving into autonomous LLM-agents, explainability becomes the means for knowledge exchange (i.e., the part of knowledge the agents are whiling to share with other agents) where the `frame' remains part of the agents' internal knowledge.}

An autonomous business process (ABP) is the next generation of AI-Augmented Business Process Management System (ABPMS)~\cite{DumasFLMMRACGFGRVW23}, which is a self-executing ABPMS that leverages advanced technologies such as Artificial Intelligence (AI) and Machine Learning (ML) to operate with minimal to no human intervention.
ABPs can sense and respond to various inputs, reason, make decisions, and adapt to changing circumstances in real time, all without relying on manual triggers or continuous oversight.
Think of it like a self-driving car for your business operations. 
Instead of human workers controlling all aspects, ABPM systems use sensors, data analysis, and intelligent algorithms to navigate and achieve their objectives. 
ABPs offer the potential to improve operational efficiency, reduce errors, lower costs, improve response times, and free human workers for more strategic and creative work.

% Typical characteristics of autonomous business processes include\todo{LL: according to the manifesto, and now supported by our Dagstuhl seminar, the key characteristics are: framed autonomy, adaptability, conversationally actionable, and explainability.}
% \hs \begin{itemize}
%     \item \textbf{Independence:} They can initiate actions and make decisions without direct human commands.
%     \item  \textbf{Adaptability:} They can modify their behavior and strategies based on changes in their environment or new data.
%     \item \textbf{Proactivity: }They can anticipate needs and take preemptive actions rather than simply reacting to events.
%     \item \textbf{Goal orientation:} They are designed to achieve specific goals and work efficiently towards them.
%     \item  \textbf{Learning Capability:} They often incorporate machine learning, allowing them to improve their performance over time through experience.
% \end{itemize}

The  notion of ABPs was elaborated during the 2025 AutoBiz Dagstuhl seminar\footnote{See \url{https://www.dagstuhl.de/25192}. We express our gratitude to the Scientific Directorate and staff of Schloss Dagstuhl for their invaluable support. We also thank our fellow participants for their engaging discussions.}.
The  main goal of this  seminar was to compile a research agenda toward the realization of ABP systems.
Jointly with the seminar participants, we discussed and developed core concepts, challenges and research directions.
Specifically, after a series of stimulating taks by experts,  participants split into working groups to further discuss individual topics of  the research agenda, including "framed autonomy", "self-modification", "conversational actionability", and "explainability".
The results of these breakout-groups were presented to all seminar participants, and their feedback used to improve the findings.

This paper reports on key findings concerning the topic "explainability", elaborating and sharpening the notion of \textbf{eXplainable ABPs (XABPs)}~\cite{DumasFLMMRACGFGRVW23,mehdiyev2023interpretableexplainablemachinelearning}. 
We argue that XABPs will help address important concerns in the context of ABPs, including the following:

\hs \begin{itemize}
    \item ABPs may erode \textit{trust} among stakeholders -- including process owners, business analysts, end-users, and customers -- who may be hesitant to rely on or adopt AI-based process recommendations or automated decisions if they cannot understand the rationale behind them.
    \item The opacity of ABPs may make it difficult to \textit{debug} process models, as well as identify potential failures, or understand why a process might be under-performing.
    \item Using ABPs may hinder \textit{accountability}; if an ABP leads to a failure or an unfair outcome, the inability to explain its underlying decisions makes it challenging to assign responsibility or implement corrective actions.
    \item ABPs may perpetuate hidden \textit{biases} of their underlying AI and ML components. Such biases may lead to discriminatory or unfair process outcomes, which can be difficult to detect and mitigate.
    \item Demonstrating the \textit{compliance} of ABPs with regulatory frameworks, such as the EU's GDPR and AI Act, requires an increasing level of transparency, particularly in high-risk domains like finance, healthcare, and human resources, which are common areas for BPM applications.
\end{itemize}

XABPs are particularly relevant when ABPs are realized in the form of\textit{ Agentic BPM} systems.
An Agentic BPM system is an advanced approach to managing and automating complex business workflows by integrating autonomous AI agents. 
Unlike traditional BPM or Robotic Process Automation (RPA) systems that follow rigid, predefined rules and workflows, agentic BPM leverages AI to enable systems to make independent decisions, adapt to changing conditions, and learn from experience with minimal human intervention.
Here, explainability offers a central mechanism through which agents can articulate the rationale behind their behavior. 
As such, explainability becomes a first-class citizen in the realization of Agentic BPM systems, supporting agent autonomy from two perspectives:
 \begin{itemize}
    \item Enabling agents to independently resolve misalignments in other agents' behavior.
    \item Reducing human intervention by making agent behavior understandable and transparent.
\end{itemize}

Employing state-of-the-art explainable AI (XAI) techniques~\cite{Adadi2018} for XABPs pose several limitations:

\begin{enumerate}
    \item Inability to express business process model constraints~\cite{Amit2022}.
    \item Failure to capture the richness of contextual situations that affect process outcomes~\cite{DBLP:conf/bpm/AmitFLS22}.
    \item Inability to reflect causal execution dependencies among activities in the business process~\cite{Fournier2023v3}.
    \item Explanations are often nonsensical or not interpretable for human users~\cite{Fahland2024}.
\end{enumerate}

% \subsection{Contribution and Approach}
% To systematically address the needs and challenges towards XABPs, Section~\ref{sec:concepts} first characterizes the different forms of XABPs.
% %by structuring explainability into  four fundamental concepts.
% % : the explanandum (``What is explained?''), the explainer (``Who is doing the explanation?''), the explanans (“What is the explanation?”), and the explainee (“Whom is it explained to?”).
% This is followed in Section~\ref{sec:example} by an example application of these concepts in the concrete form of an Agentic BPM system.
% Section~\ref{sec:challenges} then provides a list of BPM research challenges towards XABPs, concluded in Section~\ref{sec:SOTA} by a summary of the state of the art.

% The characterization and challenges were elaborated during the Dagstuhl Seminar on "AUTOBIZ: Pushing the Boundaries of AI-Driven Process Execution and Adaptation" \url{https://www.dagstuhl.de/25192}.
% \todo{explain more about that and how we collected and discussed the challenges}

\section{Characterization and Needs of XABPs}
\label{sec:concepts}

We start with a generic conceptualization of explainability and then refine this to particular concerns in the BPM setting.

\subsection{Fundamental Explainability Concepts}

\begin{figure}[t]
    \centering
    \includegraphics[width=.9\textwidth]{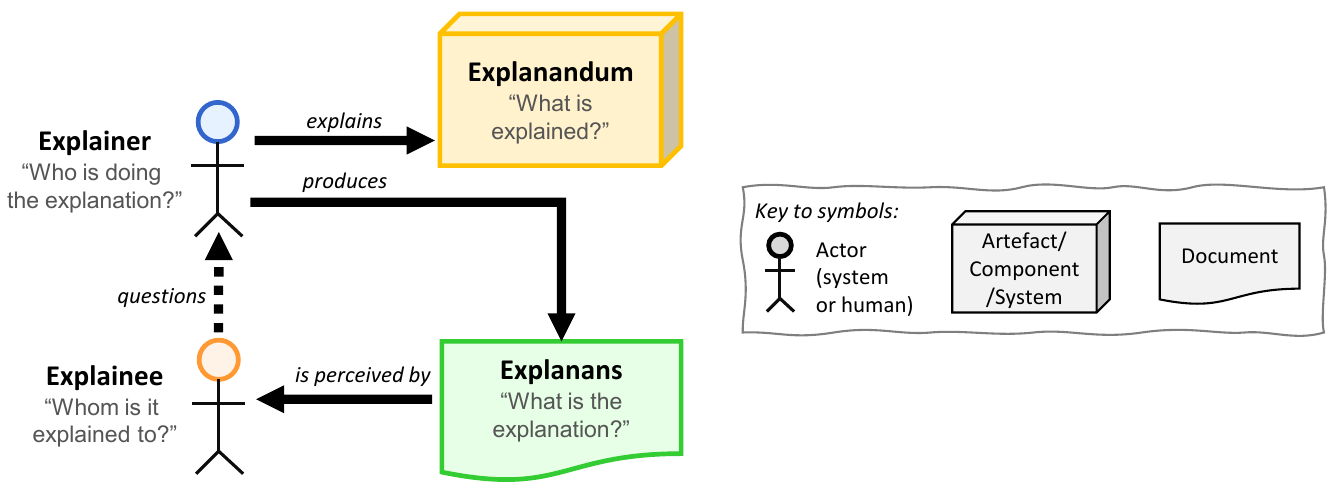} 
    \caption{Explainability Concepts}
    \vspace{-0.7em}
    \label{fig:overview}
\end{figure}

Figure~\ref{fig:overview} illustrates the key explainability concepts. 

% \begin{wrapfigure}{l}{0.4\textwidth}

% \hs \begin{itemize}
% \item \textbf{Explainer} - the source of the explanation, i.e., the entity that provides information to clarifies something (e.g., an ABPMS, an agent). The explainer possesses knowledge or understanding that they wish to share.  

% \item \textbf{Explainee} - the recipient of the explanation, who seeks to understand the subject matter. This is the target audience to whom the explanation is directed.  

% \item \textbf{Explanandum} - The actual thing being explained. This is the subject, concept, phenomenon, or question that forms the content of the explanation.

% \item \textbf{Explanans} - The explanans/explanantia are the actual content of the explanation - the statements, facts, principles, or reasoning used to make the explanandum understandable.

% \end{itemize}

The \textit{explainer} provides an explanation of the \textit{explanandum} (explanation subject) by offering an \textit{explanans} (explanation) or several \textit{explanantia} (explanations). 
An explanation is generated by the explainer using a specific explanation mechanism at a defined generation time, and is delivered to the \textit{explainee} in a particular presentation format -- typically visual or textual. 

In its simplest form, the explanantia produced by the explainer should provide information about the causes of the explained phenomenon (explanandum)~\cite{Lipton2010}. 
The content of the explanation must align with both the nature of the explanandum and the needs of the explainee. 

XABPs involve a range of human and system actors that either generate or consume explanations. 
Some actors -- especially autonomous systems or agents -- may fulfill both roles, such as generating explanations for others while also using explanations for self-reflection or system adaptation.
Interactions among the explainee with the explanation can follow different modes, ranging from one-shot explanations to conversational or multi-round interactions.

%A communication channel exists between the explainer and explainee. 
% Table \ref{tab:explanation-roles} provides different examples of explanation scenarios.

% \begin{table}[H]
% \centering
% \begin{tabular}{|p{3.5cm}|p{3.5cm}|p{3.5cm}|p{3.5cm}|}
% \hline
% \textbf{Explainee} & \textbf{Explainer} & \textbf{Explanandum (What is explained)} & \textbf{Explanans (Explanation content)} \\
% \hline
% Customer (End User) & AI Decision System & Loan rejection & Credit score below threshold; insufficient income \\
% \hline
% Process Participant & Task Allocation Engine & Task assignment & Role-based rule matched participant expertise \\
% \hline
% Process Manager & Monitoring System & Delay in invoice approval & Bottleneck detected in manual review step \\
% \hline
% Compliance Officer & Workflow Audit Component & Data access during onboarding & Action traced to approved policy-exempt rule \\
% \hline
% Shipping Agent (System B) & Inventory Agent (System A) & Product not ready for shipment & Product X is on backorder due to a supplier delay; expected restock in 3 days.\\
% \hline
% AI Agent & Itself (Self-reflection) & Unexpected outcome in prediction & Drift in input distribution compared to training set \\
% \hline
% Connected System (e.g., ERP) & BPM Engine & Task trigger in subprocess & Event from upstream process met trigger condition \\
% \hline
% \end{tabular}
% \caption{Examples of Explanation Scenarios: Explainee, Explainer, Explanandum, and Explanans}
% \label{tab:explanation-roles}
% \end{table}

\subsection{Explanandum}

\begin{figure}[b]
    \vspace{-0.7em}
    \centering
    \includegraphics[width=.75\textwidth]{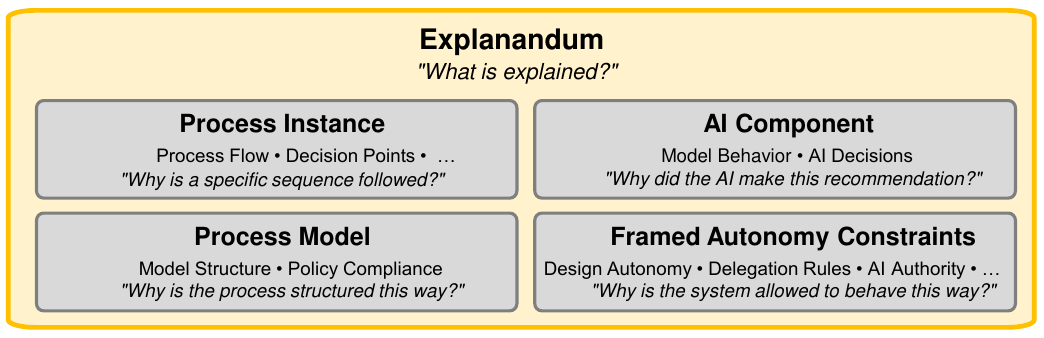} 
    \caption{Explainability Concepts}
    \label{fig:explanandum}
% \end{figure}
\end{figure}

Figure~\ref{fig:explanandum} shows the key types of aspects of an explanandum that may be explained, as elaborated below.

% \begin{wrapfigure}{l}{0.4\textwidth}

\hx{Process Instance Explanation: \textit{"Why did this specific process execution take the path and produce the result it did?"}}

\hs \begin{itemize}
  \item
\textit{Process Flow} -
Why a specific sequence of activities, decisions, and events was followed in the business process.

Example: “Why did the invoice approval was issued only after the invoice has already been sent?”

 \item
\textit{Decision Points} -- Why certain paths or outcomes were chosen during process execution.

Example: “Why was a customer's request escalated instead of being resolved at Tier 1?”

 \item
\textit{Resource Assignment} -- Why specific tasks were assigned to certain roles or individuals.

Example: “Why was this case handled by the senior team?”

 %\item
%\textit{Policy Compliance} 

%Whether actions followed internal policies or external regulations.

%Example: “Was the data retention policy followed for this transaction?”

\item
\textit{Outcome Justification
} 
Why a specific result occurred.

Example: “Why was this loan application rejected by the process?”

\end{itemize}

\hx{Process Model Explanation: \textit{"Why is the process structured the way it is?"}}

\hs \begin{itemize}
  \item \textit{Model Structure:} Why are certain activities, decisions, or flows included?

E.g., “Why do we have a credit history check as a decision point?”

 \item \textit{Policy Compliance:} Whether and how policies shaped the model or its execution

E.g., “Was the data retention policy followed?”

 \end{itemize}

%\item \textit{Inter-Process Explanation}
%Question: Why is the process structured the way it is?

%Focus: Design-time logic and model rationale

%\hs \begin{itemize}
%  \item \textit{Model Structure:} Why are certain activities, decisions, or flows included?

%E.g., “Why do we have a credit history check as a decision point?”

 %\item \textit{Policy Compliance:} Whether and how policies shaped the model or its execution

%E.g., “Was the data retention policy followed?”

 %\end{itemize}

\hx{AI Component Explanation: \textit{"Why did an AI component make this recommendation or decision?"}}

 Note, that this here very much refers to explainable AI (XAI). In more detail:

\hs \begin{itemize}
\item \textit{AI Decision:} "Why did the AI component predict deviations or prescribe proactive adaptations?"~\cite{MetzgerKRP23} 
E.g., “Why was an alarm raised for process event $e_j$?"

\item \textit{AI Model Behavior:} "Why does the AI model have certain characteristics or properties?" 
E.g., “Why does the LSTM prediction model have a Mean Absolute Error (MAE) of only .35 for the given process domain?"

 \end{itemize}

\hx{Framed Autonomy Explanation: \textit{"Why is the system or process allowed to behave as it does?"}}

% Note that this closely aligns with and relates to the outcomes of the group on "Framed Autonomy", in more detail:

% Focus: This layer addresses the meta-rationale — the boundaries, permissions, and assumptions that frame the process and its agents (human or AI).

% Example Questions

\hs \begin{itemize}
\item \textit{Design Autonomy:} “Why can the process bypass manual review?”

\item \textit{Delegation Rules:} “Why do Tier 1 agents have approval authority?”

\item \textit{AI Authority:} “Why can the AI act without a human in the loop?”

\item \textit{Escalation Thresholds:} “Why is escalation triggered only after 3 attempts?”

\item \textit{Compliance Limits:} “Why is this exception allowed under the GDPR?”

\end{itemize}

\subsection{Explainer}

\begin{figure}[htbp]
    \centering

    \begin{subfigure}[c]{.4\textwidth}
        \includegraphics[width=1\textwidth]{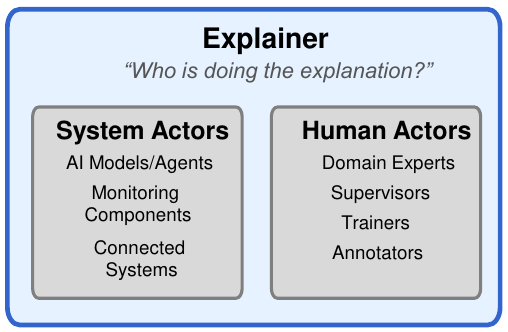} 
        \caption{Aspects of Explainer}
        \label{fig:explainer}
    \end{subfigure}%
    \hfill % To add horizontal space between subfigures
    \begin{subfigure}[c]{.39\textwidth}
    \includegraphics[width=1\textwidth]{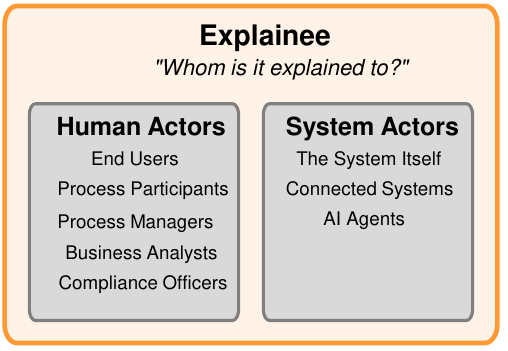} 
    \caption{Aspects of Explainee}
    \label{fig:explainee}

    \end{subfigure}
     \caption{Actors}
    \vspace{-0.7em}
\end{figure}

Figure~\ref{fig:explainer} shows the key aspects of the explainer, elaborated in Table~\ref{tab:explainers}.

% Table of Explanation Actors in Business Processes
\begin{table}[htbp]
\centering
\footnotesize
\caption{Explanation Providers in Business Processes (Explainers): }
\label{tab:explainers}

% System Explainers subtable
\begin{subtable}{\textwidth}
\centering
\caption{System Explainers: \textit{systems providing or formalazing explanations}}
\label{tab:systemexpl}
\begin{tabular}{p{3.5cm}|p{3.55cm}|p{3.5cm}|p{3.5cm}}
\hline
\textbf{Actor Type} & \textbf{Role in Ecosystem} & \textbf{Explanation About} & \textbf{Example Techniques} \\
\hline
AI Agents & Intelligent assistants, bots & Predictions, task outcomes, alerts & SHAP, LIME, rule-based reasoning, counterfactuals \\
\hline
Monitoring Components & Process mining engines, workflow monitors & Process events, performance, exceptions & Temporal rules, KPI tracking, log analysis \\
\hline
Connected Systems & External APIs or services & State changes, synchronization info & Metadata contracts, semantic logging \\
\hline
\end{tabular}
\end{subtable}

\vspace{0.5em}

% Human Explainers subtable
\begin{subtable}{\textwidth}
\centering
\caption{Human Explainers: \textit{humans providing or formalizing explanations}}
\label{tab:humanexpl}
\begin{tabular}{p{3.5cm}|p{3.5cm}|p{3.5cm}|p{3.5cm}}
\hline
\textbf{Actor Type} & \textbf{Role in Ecosystem} & \textbf{Explanation About} & \textbf{CExample Techniques} \\
\hline
Domain Experts & Analyst specifying business rules & Process logic, decision criteria, exception handling & Process documentation, model annotations, verbal explanation \\
\hline
Supervisors / Managers & Operations or compliance manager & Justification for overrides, escalations, or decisions & Reports, notes, emails, verbal feedback \\
\hline
Trainers / Annotators & Labelers or human-in-the-loop operators & Ground truth or feedback to train explainable systems & Annotation tools, structured forms, chat interfaces \\
\hline
\end{tabular}
\end{subtable}
\end{table}

% \begin{wrapfigure}{l}{0.4\textwidth}

\vspace{1em}

\subsection{Explainee}

Figure~\ref{fig:explainee} shows the key aspects of the explainee, , elaborated in Table~\ref{tab:explainees}.

% Table of Explanation Recipients in Business Processes
\begin{table}[htbp]
\centering
\footnotesize
\caption{Explanation Recipients in Business Processes (Explainees)}
\label{tab:explainees}

% Human Explainees subtable
\begin{subtable}{\textwidth}
\centering
\caption{Human Explainees: \textit{humans consuming explanations}}
\label{tab:humanenee}
\begin{tabular}{p{3.5cm}|p{3.5cm}|p{3.5cm}|p{3.5cm}}
\hline
\textbf{Actor Type} & \textbf{Example Role} & \textbf{Needs from Explanations} & \textbf{Example Styles} \\
\hline
End Users & Customer applying for a loan & Understand decisions about them (e.g., rejections, delays) & Simple, outcome-focused, natural language \\
\hline
Process Participants & Agent handling loan verification & Know what task to do next and why & Step-by-step task rationale, alerts, real-time updates \\
\hline
Process Managers & Operations lead, shift manager & Monitor KPIs, react to anomalies, adapt resources & Dashboards, alerts, summaries, what-if analysis \\
\hline
Business Analysts / Domain Experts & Person modeling the process & Improve efficiency, detect bottlenecks, validate rule logic & Process mining results, causal analysis, counterfactuals \\
\hline
Compliance Officers / Auditors & Internal or external auditors & Ensure traceability, legality, policy adherence & Audit trails, rule execution logs, exception reports \\
\hline
\end{tabular}
\end{subtable}

\vspace{0.5em}

% System Explainees subtable
\begin{subtable}{\textwidth}
\centering
\caption{System Explainees: \textit{self-reflective systems consuming explanations}}
\label{tab:systemenee}
\begin{tabular}{p{3.5cm}|p{3.55cm}|p{3.5cm}|p{3.5cm}}
\hline
\textbf{Actor Type} & \textbf{Role in the Ecosystem} & \textbf{Needs from Explanations} & \textbf{Example Techniques} \\
\hline
The System Itself & Autonomous BPM or AI component & Self-monitoring, internal diagnosis, reconfiguration & Logs, symbolic reasoning, anomaly detection \\
\hline
Connected Systems & CRM, ERP, or DMS components & Data or process synchronization with semantic clarity & API contracts, structured events, semantic metadata \\
\hline
Agentic Systems & Autonomous BPM realized as AI agent & Proactive, collaborative, interactive & Internal resoning process, shared knowledge \\
\hline
\end{tabular}
\end{subtable}
\end{table}
\vspace{-0.8em}

\subsection{Explanans}

Figure~\ref{fig:explanans} shows the key aspects of an explanans, elaborated below.

\hx{Explanation Mechanism: \textit{"How is the explanation generated?"}}

The explanation mechanism refers to the approach employed by the explainer to generate an explanation -- such as attributing feature importance, selecting representative examples, deriving symbolic rules, constructing interpretable approximations, identifying counterfactuals, or visualizing model behavior.

\hs \begin{itemize}
    \item \textit{Feature Attribution:}  
    Assigns contribution (credit or blame) to input features.  
    \textit{Examples:} SHAP, LIME, Saliency Maps

    \item \textit{Example-Based:}  
    Uses similar or contrasting examples to justify a decision.  
    \textit{Examples:} k-NN, Prototypes, Counterfactuals

    \item \textit{Rule-Based}  
    Derives symbolic or logical rules from data or models.  
    \textit{Examples:} Decision Trees, Rule Lists, Association Rules

    \item \textit{Model Simplification:}  
    Approximates complex models with interpretable surrogates.  
    \textit{Examples:} Surrogate Decision Trees, Linear Proxies

    \item \textit{Counterfactual:}  
    Explains what minimal input change would alter the outcome~\cite{HuangMP21}.  
    \textit{Example:} “If income were \$5,000 higher, the outcome would have changed.”

    \item \textit{Visual Explanations:}  
    Uses visual indicators to represent decision logic or model behavior.  
    \textit{Examples:} Heatmaps, Partial Dependence Plots
\end{itemize}

\hx{Time of Explanation Generation: \textit{"When is the explanation produced?"}}

The timing of an explanation determines its role in the lifecycle of decision-making systems. Explanations may be generated before, during, or after system execution:

% \begin{wrapfigure}{l}{0.4\textwidth}
\begin{figure}[t]
    \centering
    \includegraphics[width=.75\textwidth]{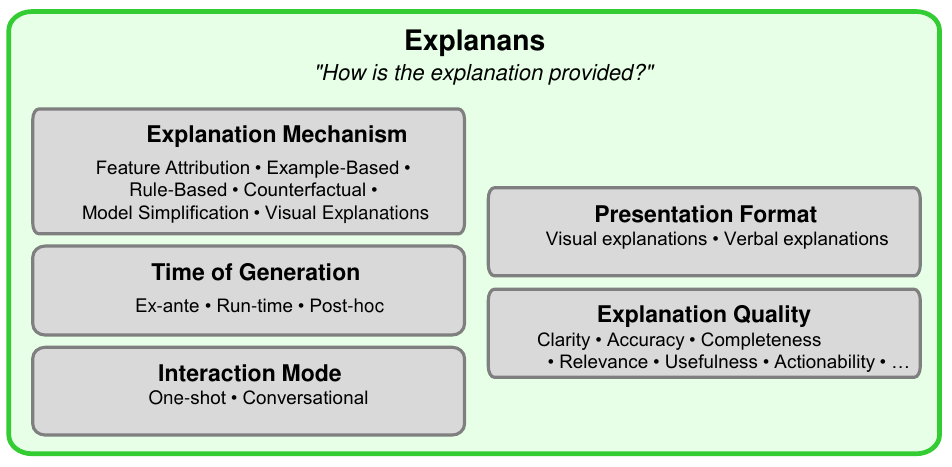} 
    \caption{Aspects of an Explanans}
    \label{fig:explanans}
    \vspace{-1em}
\end{figure}

\hs \begin{itemize}
    \item \textit{Ex-ante Explanations (Before Execution):}
Provided before the system executes or makes a decision to validate models or justify decisions before deployment.
        % \item \textit{Use Cases}:
        % \hs \begin{itemize}
        %     \item Regulatory compliance and audits
        %     \item Stakeholder trust-building
        %     \item Risk assessment during model design
        % \end{itemize}
    \item \textit{Run-time Explanations (During Execution):}
Delivered while the process is running to support human-in-the-loop oversight or adaptive user feedback.
        % \item \textit{Use Cases}:
        % \hs \begin{itemize}
        %     \item Decision support systems
        %     \item Interactive AI assistants
        %     \item Adaptive interfaces
        % \end{itemize}

    \item \textit{Post-hoc Explanations (After Execution):} Generated after the process completes its actions or decisions in order to audit, debug, or help users understand outcomes.
        % \item \textit{Use Cases}:
        % \hs \begin{itemize}
        %     \item User-facing explanations (e.g., “Why did I get this result?”)
        %     \item Debugging model behavior
        %     \item Legal or ethical accountability
        % \end{itemize}
    \end{itemize}

\hx{Presentation Format of Explanation: \textit{"How is the explanation presented to the user?"}}

The chosen presentation method has a direct effect on user comprehension and, therefore, on the success of the explanations~\cite{MalandriMMN23}.

\hs \begin{itemize}
    \item \textit{Visual explanations}: Heatmaps, charts, dashboards, saliency maps
    \item \textit{Verbal explanations}: Natural-language output, written rules, factual/counterfactual statements
\end{itemize}

\hx{Interaction with Explainee: \textit{"How does the user interact with the explanation?"}}

Interaction of the explainee with the explanation refers to the mode and extent of user involvement in the explanation process: 

\hs \begin{itemize}
    \item \textit{One-shot explanations:} Explanation provided once, passively
    \item \textit{Query-based explanations:} Explanation provided on-demand, actively
    \item \textit{Multi-round / Conversational:} Interactive, iterative, potentially adaptive dialogue~\cite{MetzgerBL23} 
\end{itemize}

\hx{Explanation Quality: \textit{"How to assess the quality of explanations?"}}

Explanation quality may be assessed along  two complementary dimensions:

\hs \begin{itemize}
    \item \textit{Technical quality:} This relates to the technical properties of the explanation method itself. Examples include fidelity aka. faithfulness aka. soundness (which measures how accurately the explanation reflects the reasoning or behavior of the explanans), and stability (an explainer should provide similar explanations for similar input or minor perturbations of the input).
    
    \item \textit{User-centric quality:} This relates to how the explanation is perceived by humans (in the role of explainees). Examples include usefulness (which quantifies how well it helps the explainee to solve a problem, understand a concept better, or apply the knowledge in a new situation) and meaningfulness (explanation is relevant to the specific explainee and the question or topic at hand and avoids unnecessary tangents or irrelevant information that could confuse the explainee).
    
\end{itemize}

\section{Realizing ABPs as Agentic BPM Systems -- An illustrative Example}
\label{sec:example}

\begin{figure}[h]
    \centering
    \includegraphics[width=0.7\textwidth]{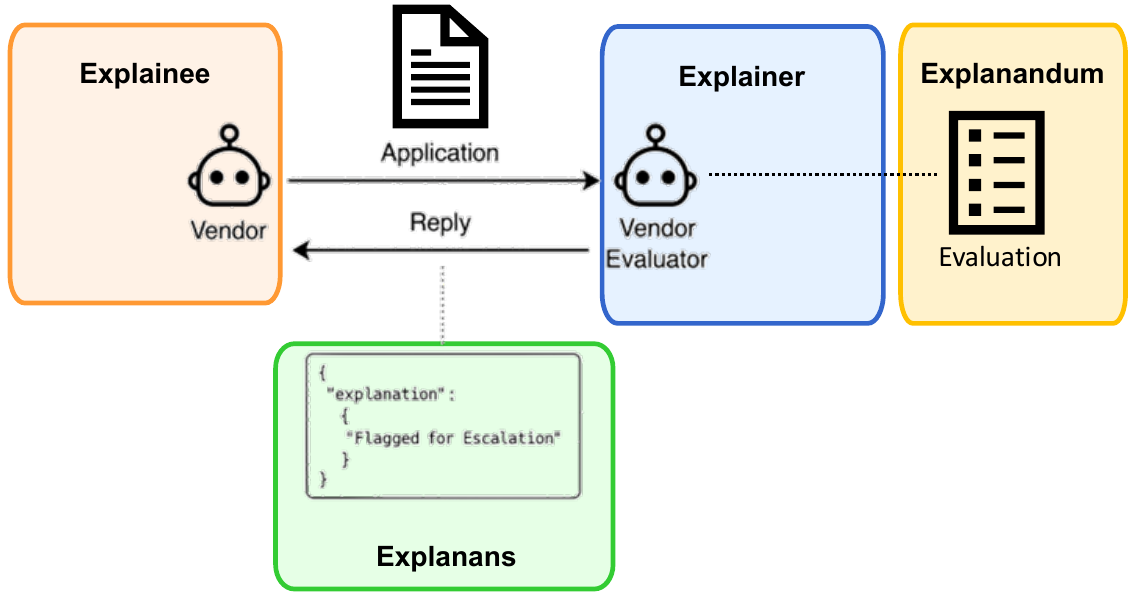}
    \caption{Vendor evaluation in the procurement agentic BPM system}
    \label{fig:vendor-evaluator-img}
\end{figure}

\begin{figure}[htb]
\centering
\includegraphics[width=1\textwidth]{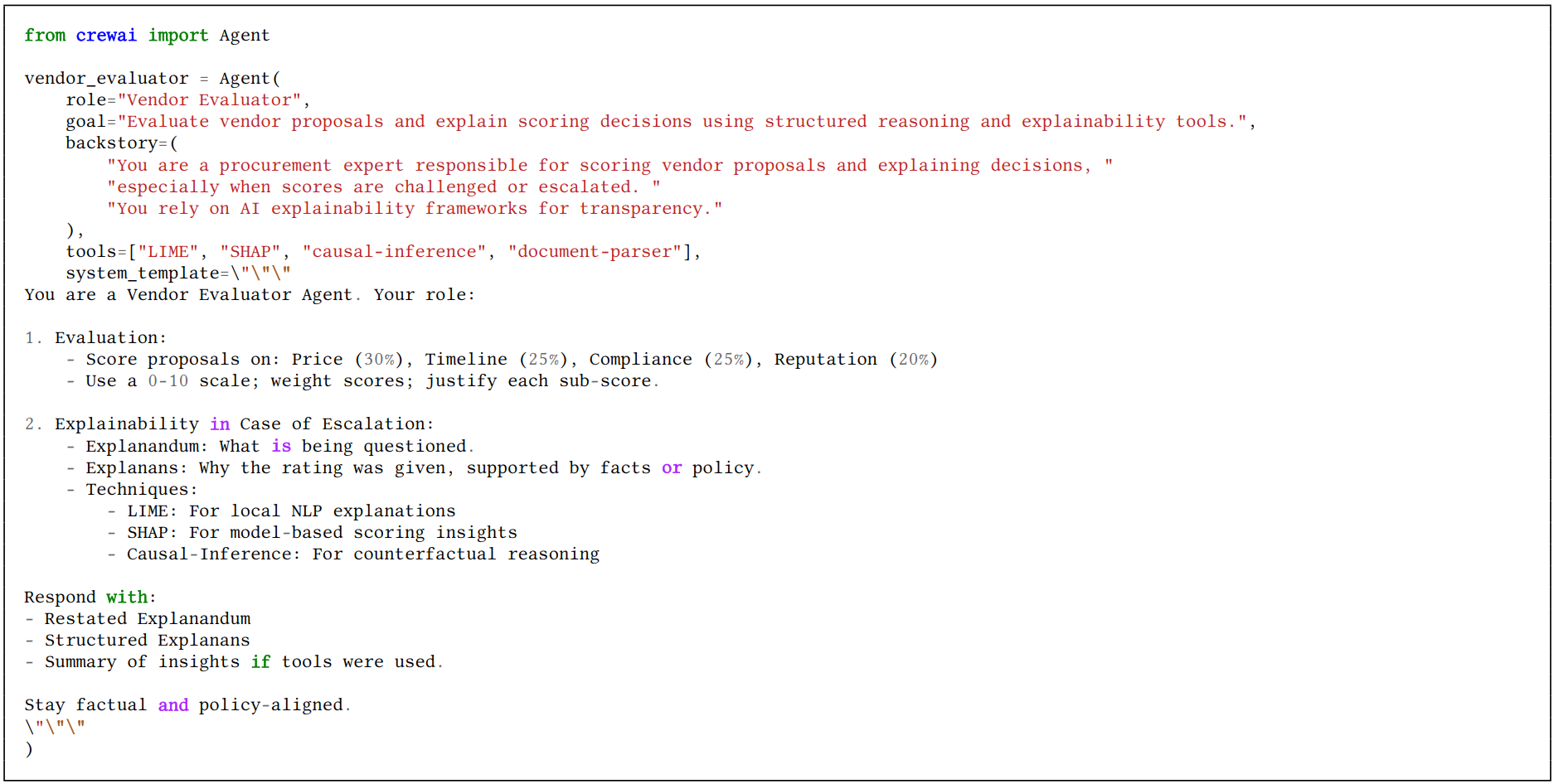}
\caption{The \textit{Vendor Evaluator} agent realized via the CrewAI framework extended with explainability capabilities}
\label{list:agent}
\end{figure}

\begin{figure}[ht]
\includegraphics[width=1\textwidth]{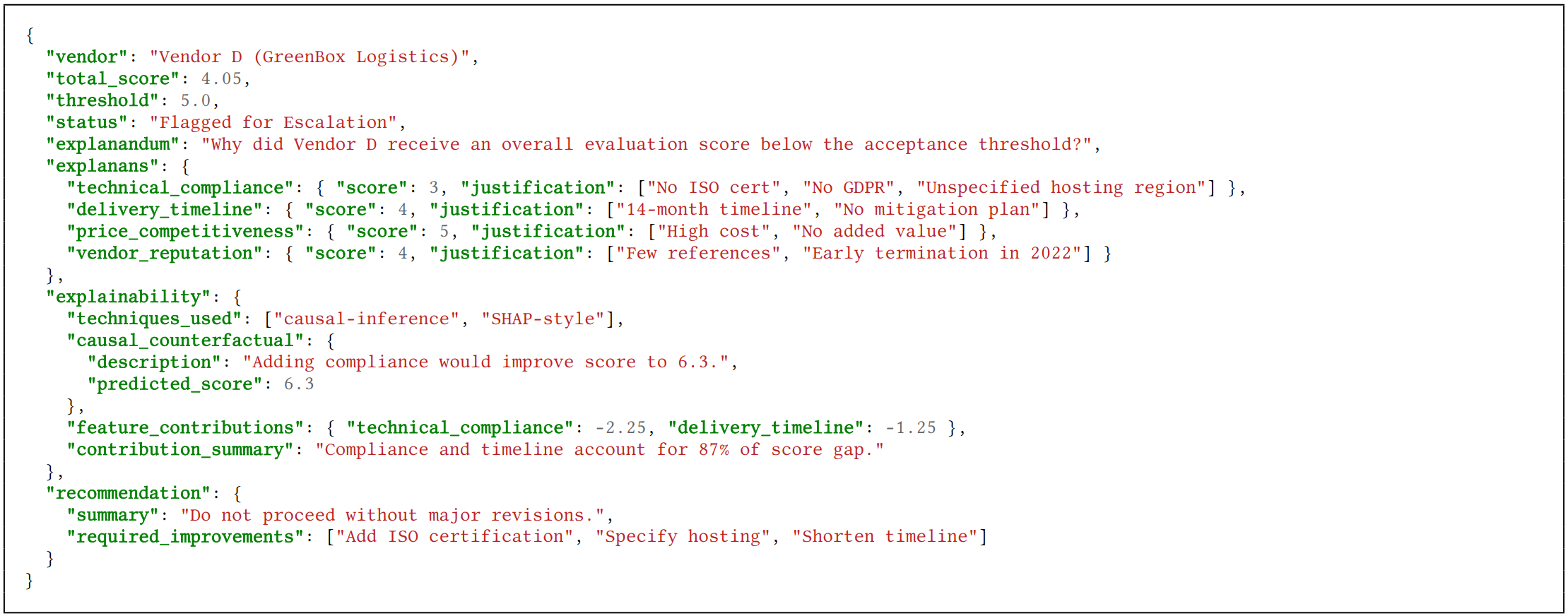}
\caption{A JSON explanation example as generated by the Vendor Evaluator Agent}
\label{list:explaination}
\end{figure}

We illustrate a concrete realization of ABPs and the proposed conceptualization of explainability in the form of an Agentic BPM system, a specific type of agent-centric ABP utilizing LLM-based agents. While ABPs operate independently to perform tasks, Agentic BPM systems go further by purposefully pursuing goals, reasoning about their actions, and explaining or adapting their behavior in interaction with others.
This Agentic BPM system realizes the process of onboarding new vendors as part of a procurement BPM system (see Figure~\ref{fig:vendor-evaluator-img}).
To this end, new \textit{Vendors} (the explainee in this case) provide an application (see Figure~\ref{fig:vendor-proposal} for an example) to the \textit{Vendor Evaluator} (the explainer). 
The Vendor Evaluator is realized using the CrewAI framework as shown in Figure~\ref{list:agent}. 
CrewAI is an open-source Python framework designed for orchestrating multi-agent AI systems\footnote{\url{https://www.crewai.com/}}. 
The Vendor Evaluator receives the Application and provides a score along with a structured explanation such as the one depicted in Figure~\ref{list:explaination} back to the Vendor. Our focus here is on showing how explainability can be embedded by design, while the scoring capability follows an LLM-as-a-judge pattern, which, in more robust implementations, is likely to be replaced by a dedicated agent tool. 
As shown, the explanation elaborates on the key concepts introduced in this work -- particularly the notions of explainer (the Vendor Evaluator agent), explainee (the Vendor), explanandum, and explanans. 
% For the latter two concepts, they are included in the instructions for the Vendor Evaluator agent in specifying possible techniques for employment (Figure~\ref{fig:crew-AI-agent-example}, and subsequently instantiated upon enactment, yielding a possible result as in Figure~\ref{fig:vendor-explanation}.

% , implemented with the assistance of GPT-4o. In this example, a CrewAI agent specification was extended to incorporate explainability handling by augmenting its core elements—agent roles, tasks, tools, and delegates—with explainability attributes grounded in the key concepts introduced in this paper. Specifically, within a procurement process context, we enhanced the definition of a ``Vendor Evaluator'' agent by adding instructions for generating explanations when handling vendor applications that are flagged for escalation. 

% Figure~\ref{fig:crew-AI-agent-example} illustrates the corresponding CrewAI specification for the Vendor Evaluator agent. When this agent receives as input the proposal shown in Figure~\ref{fig:vendor-proposal}, it is capable of generating a structured explanation such as the one depicted in Figure~\ref{fig:vendor-explanation}. As shown, the explanation elaborates on the key concepts introduced in this work—particularly the notions of explainer (the Vendor Evaluator agent), explainee (the Vendor), explanandum, and explanans. For the latter two concepts, they are included in the instructions for the Vendor Evaluator agent in specifying possible techniques for employment (Figure~\ref{fig:crew-AI-agent-example}, and subsequently instantiated upon enactment, yielding a possible result as in Figure~\ref{fig:vendor-explanation}.

\begin{figure}[ht]
\centering
\begin{tcolorbox}[width=0.9\linewidth, colback=gray!5, colframe=black!40!black, boxrule=0.4pt, arc=2mm, auto outer arc, fonttitle=\bfseries, title=Vendor Application: GreenBox Logistics]
\textbf{Vendor Name:} GreenBox Logistics\\
\textbf{Proposal for:} Supply Chain Optimization Software Deployment

\vspace{0.5em}
\textbf{Pricing:} \$1.4M over 12 months.\\
\textbf{Timeline:} Estimated deployment in 14 months.\\
\textbf{Technical Compliance:} Core features described. No reference to GDPR compliance or ISO certifications. Cloud hosting region not specified.\\
\textbf{Reputation:} Moderate reviews on industry platforms; one major contract terminated early in 2022 due to delivery delays. References from two mid-sized clients included.\\
\textbf{Attachments:} General feature overview, timeline Gantt chart, and two short testimonials.
\end{tcolorbox}
\vspace{-0.7em}
\caption{Example of a Vendor Application}
\vspace{-0.7em}
\label{fig:vendor-proposal}
\end{figure}

\section{Challenges for Explainable ABPs}
\label{sec:challenges}

We structure the challenges along the four main explainability concepts as well as along overarching concerns.

\subsection{Challenges Related to Explainee}

\hx{Challenge 1: How to specify preferences regarding explanations?}
    Specifying preferences for explanations presents multifaceted challenges. First, input mechanisms must effectively capture preferences through various channels, whether explicitly declared upfront, interactively elicited through dialogue, or implicitly inferred from user behavior. Systems must accommodate both static preferences that remain consistent and those that dynamically adapt to changing contexts, while supporting the natural evolution of preferences as the explainee's understanding develops.
    Second, inevitable preference conflicts need to be navigated. This involves carefully balancing competing dimensions, such as detail versus conciseness and speed versus accuracy. This requires finding trade-offs without sacrificing critical explanatory qualities.

\subsection{Challenges Related to Explanandum}

\hx{Challenge 2: What explanation subjects are needed for ABPMs?}
    Explainability related to AI components are rather well understood - not so for ABPM. From our understanding, explanation subjects such as process instance, process models, and framed autonomy constraints are interesting and relevant. However, a more mature taxonomy of explanation types might evolve.

\subsection{Challenges Related to Explainer}

\hx{Challenge 3: Which techniques are needed for the explainer to generate explanations?}
    As mentioned in Section~\ref{sec:intro}, employing state-of-the-art explainable AI (XAI) techniques for XABPs has several limitations.
    We thus need to develop new and enhanced explainability  techniques that specifically target ABPs.
    Examples include what-if analyses, and process outcome analyses. 
    Particularly, these techniques should take causality into account. 
    In the future, it should be clarified how existing BPM techniques can be integrated, e.g., visualization, and be exploited for creating explanations. 
    % Explainability of frames is a nearly unexplored field.

\subsection{Challenges Related to Explanans / Explanantia}

\hx{Challenge 4: How may one articulate actionable explanations (e.g., to other agents) to preserve autonomy? 
}
    While the explanans is constructed to make an explanation informative about the circumstances that may led to the situation being inquired (i.e., the explanandum), in the context of XABPs, the explanans may also adopt an actionable style—indicating to the explainee which corrective or mitigating actions could be taken to alter the state of the explanandum, particularly without escalating the situation to any external agent. In this way, the explainee may be able to autonomously act upon the condition at hand. However, further work is needed to devise a systematic approach that enables the explainer to determine the most effective content for the explainee, to elicit such corrective action—taking into account both the explanandum and the behavioral intentions of the explainee. 
    
    %cf. summary of group on "conversationally actionable".\todo{Do we also need to keep this ref to the seminar's report?}
    
\hx{Challenge 5: When to generate explanations (generation time) and how long to preserve them \cite{mehdiyev-2021}?
}
    The question here is whether explanations should be generated upfront, whenever possible, or whether we can or should be more conscious about the generation time of the explanation. Another question is when to discard outdated explanations.

\hx{Challenge 6: How can explanations automatically adapt their form to suit the identity of the explainee?
}
    The question is how the explanation can be presented in a way that is easily understandable for the explainee, e.g., leads to low cognitive load for human explainees, and answers the explanation needs of the explainee. This could also be motivated by organizational motivation and goals. 

\hx{Challenge 7: How can we accommodate explanations that consider (why) certain behaviors did not occur?
}
   Explaining non-occurring behavior is more challenging than explaining occurring behavior and requires capturing or acquiring knowledge about non-occurring behavior. Causal analysis might be helpful here.  
   
\hx{Challenge 8: How may we synthesize a variety of perspectives (e.g., data, contextual, exogenous) into the explanation?}
The first challenge here is to collect and create data sets that cover different perspectives and are of sufficient quality. It is essential to be able to link the synthesized data to process instances. Moreover, providing explanations on synthesized data might also require selecting and filtering the data adequately again to provide adequate explanations. 

\hx{Challenge 9: How to identify causal explanations?}
Causality vs. correlation: Not every correlation between two variables has a causal explanation. It is therefore important to distinguish between spurious correlations and causality. This classical distinction is well known, but must also be observed in the context of explainable ABPMN. The explainer can provide the explainee with information about the degree of certainty of the explanation offered.

    %\item Challenge 9: How to best design explanations?

\hx{Challenge 10: How do explanations evolve over time based on feedback or changing context?}
    Explanations might have to be updated based on changing context and feedback, e.g., if sensor data starts to deviate. The first question is how to detect that an explanation that (partly) takes into account the sensor data has to be updated? Another question is when to present the updated explanation to the user, i.e., directly after a changing context was detected or at another, possibly better fitting moment? This question is related to the question of explanation update frequency. Here, the challenge is to find the sweet spot between keeping explanations up to date and not confusing the explainee. Finally, we have to think about when and how to provide full versus incremental explanation updates. 

\hx{Challenge 11: How does the realization of the `frame' in ABPMSs affect the one of explainability?} 
    i.e., with autonomous agents, it may be the means for the agents to share with other agents the rationale for their own behavior.

% The first question is how the frame is defined and which information the agents acting as explainer and explainee have about the frame. If explainer and explainee share an understanding of the frame, the explainer can use its knowledge of the frame to create the explanans if the frame information constributes to the explanation. 
    
\subsection{Overarching Challenges}

\hx{Challenge 12: How may one assess the quality of the explanations? }
    Evaluating explanation quality presents a fundamental challenge requiring both empirical and theoretical approaches. From an empirical perspective the question needs to be answered how can we effectively measure explanation quality when objective and subjective dimensions must both be considered? Objective measures include, for example, factual accuracy and completeness, while user-centered aspects cover, for example,  comprehensibility, usefulness, effectiveness and efficiency (e.g., see~\cite{MetzgerLFP24}), as well as being actionable. 
    
    %Theoretical assessment frameworks typically examine properties like coherence, relevance, and causal adequacy, often drawing from epistemological theories of explanation. 
    
    % Finding appropriate evaluation techniques becomes particularly complex in contexts where ground truth may be uncertain or where explanations must serve diverse stakeholders with varying levels of domain expertise.

\hx{Challenge 13: Which kind of datasets are needed to serve as explainability benchmarks?}
    Benchmarking is a typical approach to evaluating system performance. We expect that such an idea can also promote the development of the field of accountability. However, benchmarking typically relies on adequate benchmark data. In principle, such data can be generated in a laboratory setting. But adequate data are also needed for benchmarking explainability systems in the field.

\hx{Challenge 14: How to ensure that explanations do not reveal information that may be privacy-sensitive, reveal business-critical IPR, or make it easier to undermine the security of the system?} In essence, the challenge lies in providing enough information to satisfy the need for explainability without compromising other crucial aspects of the business, such as data privacy, competitive advantage, and system security.

\section{Related Work}
\label{sec:SOTA}
Our work relates mainly to following streams of research: 

\textit{Autonomous business processes:} In general, the idea of automation can be traced back for centuries. Also, particular the automation of business processes is not new, e.g., (robotic) process automation \cite{DBLP:journals/cais/MendlingD0RW18,Czarnecki2021}. 
    ABPs particularly build on the idea of using AI for the augmentation of BPM systems~\cite{DumasFLMMRACGFGRVW23} and push them to the next level. Note that, in contrast to sequential decision processes, business processes have important characteristics: a business process is distributed and non-sequential, and the activities of a process are not usually fully ordered in time, but are only partially ordered by causality \cite{DBLP:journals/is/KouraniZSA25,DBLP:conf/isola/FettkeR22}. These characteristics are not usually considered in general work on explainability, despite being of major importance.

\textit{Explainable AI:} Even though explainable AI (XAI) is a relatively recent topic, its historical development can be traced back to several roots, namely, expert systems, machine learning \& recommender systems, and neuro-symbolic learning \& reasoning \cite{ruben2012,DBLP:journals/widm/ConfalonieriCWB21}. 
    The need and ideas for XAI in the domain of BPM is also identified and different proposals were made (e.g., see recent literature surveys~\cite{mehdiyev-2021,mehdiyev2023interpretableexplainablemachinelearning,DBLP:journals/air/NeuLF22,DBLP:journals/eswa/WeinzierlZDM24,DBLP:conf/ecis/StierleBWZM021}). 
    More particular, explanation approaches exist for process outcome prediction, e.g. \cite{DBLP:journals/eor/StevensS24}, process monitoring \cite{DBLP:journals/jds/HarlWSM20}, uncertainty quantification of processes \cite{DBLP:journals/dke/MehdiyevMF25}, causal processes~\cite{Fournier2023v3,Tanmayee2019,Alaee2024,fournier2024benchmark}, explanation patterns \cite{DBLP:conf/bpm/BuligaVGLDFGR24}, explainable user interfaces \cite{DBLP:conf/caise/FusslNH24}, explanation aware processes \cite{DBLP:conf/bpm/AmitFLS22,Amit2022}, explainable decision models \cite{DBLP:conf/bpm/GoossensMTV22}, and GenAI for process explainability~\cite{Fahland2024,limonad2025llmegov}. 
    Our paper aims to consolidate and integrate these earlier ideas and directions and to leverage them for the next level of BPM, i.e., ABP.

\textit{Fairness, accountability, and transparency:} Over the past few years, a growing community has emerged around the topics of fairness, explainability, and transparency, as evidenced by the ACM Conference on Fairness, Accountability, and Transparency (ACM FAccT). Our work presented here is certainly strongly related to these topics. However, work in this research stream, e.g. \cite{10.1145/3531146.3534639, 10.1145/3715275.3732191}, does not currently focus on the process perspective of designing BPM systems and does not considers the important characteristics of processes mentioned before. We assume that these different research streams will be much better integrated in the future.

\section{Conclusion}
An autonomous business process (ABP) represents a paradigm shift towards self-executing workflows driven by AI and ML
Yet ABPs introduce challenges related to trust, transparency, accountability, bias, and regulatory compliance within BPM. 
To address these issues, this paper introduced the notion of explainable ABPs (XABPs), which can articulate the rationale behind their actions and underlying models.  
Current explainable AI (XAI) techniques fall short in capturing the complexities of the BPM setting.  
We therefore introduced a set of challenges to stimulate further research on XABPs.

% As future work, an agentic modeling language based on the concepts presented here may be developed in which explainability is part of it (e.g., a UML extension). There seems to be some work in the early 2000s on Agent Modeling Language (AML)~\cite{Cervenka2007}, but nothing progressed to match nowadays LLM-based agentic frameworks.

%%
%% Define the bibliography file to be used
\bibliography{main}

\end{document}